\documentclass[twocolumn,showpacs,preprintnumbers,amsmath,amssymb,prb]{revtex4}

\usepackage{graphicx}
\usepackage{dcolumn}
\usepackage{bm}

\begin{document}

\preprint{APS/123-QED}

\title{Reply to ``Comment on `Transition from Bose glass to a condensate of triplons in Tl$_{1-x}$K$_x$CuCl$_3$'\,''}

\author{Fumiko Yamada$^1$}
\author{Hidekazu Tanaka$^1$} 
\author{Toshio Ono$^1$}
\author{Hiroyuki Nojiri$^2$}

\affiliation{
$^1$Department of Physics, Tokyo Institute of Technology, Meguro-ku, Tokyo 152-8551, Japan\\
$^2$Institute for Material Research, Tohoku University, Aoba-ku, Sendai 980-8577, Japan
}

\date{\today}

\begin{abstract}
Showing low-temperature specific heat and other experimental data and also on the basis of established physics, we argue against the comment made by Zheludev and H\"{u}vonnen criticizing our recent study on the magnetic-field-induced spin ordering and critical behavior in Tl$_{1-x}$K$_x$CuCl$_3$, which is described as the Bose glass-condensate transition of triplons.
\end{abstract}

\maketitle


In our recent paper~\cite{Yamada} referred to as paper I, we reported specific heat and ESR studies of the magnetic-field-induced Bose glass to Bose-Einstein condensate transition of triplons and the critical behavior in Tl$_{1-x}$K$_x$CuCl$_3$. However, Zheludev and H\"{u}vonnen~\cite{Zheludev} criticize that the specific heat peak shown in Fig.~1(b) in paper I is rounded; thus, the ambiguity of the transition field $H_{\rm N}$ is ${\pm}\,0.2$\,T at low temperatures and leads to the different critical exponents ${\phi}\,{=}\,1.74$ and 1.39 for $x\,{=}\,0$ and 0.36, respectively. Here, the critical exponent ${\phi}$ is defined as
\begin{eqnarray}
H_{\rm N}(T)-H_{\rm c}\,{=}\,AT^{\phi}, 
\label{eq:powerlaw}
\end{eqnarray}
where $H_{\rm N}(T)$ and $H_{\rm c}$ are the transition fields at finite and zero temperatures, respectively. They also criticize that the specific heat in Tl$_{1-x}$K$_x$CuCl$_3$ does not diverge at $H_{\rm N}$; thus, the transition is not a continuous thermodynamic transition but a crossover. They ascribe the absence of the true phase transition to the staggered $g$-tensor and the Dzyaloshinsky-Moriya (DM) antisymmetric interaction. In what follows, we argue against their criticisms.

Actually, the low-temperature specific heat data shown in Fig.~1(b) in paper I appear shrunk, because we plotted many field scan data measured at various temperatures. In Fig.~1, we show the increase in the specific heat vs magnetic field measured at 0.45\,K. Although the specific heat peak is rounded, we can determine the transition field $H_{\rm N}$ within an error of ${\pm}\,0.05$\,T, which is approximately the same as the size of symbols in Figs.~1(c) and 2 in paper I. The error in determining $H(T)_{\rm N}$ is not as large as ${\pm}\,0.2$\,T. 

For $x\,{\neq}\,0$, the transition field $H(T)_{\rm N}$ increases rapidly with temperature. The effect of the error on the estimation of the critical exponent is small. As shown in Fig.~2 in paper I, the low-temperature phase boundary for $x\,{\neq}\,0$ cannot be described by a single exponent $\phi$, although Comment authors claim that the phase boundary can be expressed by the single exponent ${\phi}\,{=}\,1.39$. To investigate the change in the exponent with a fitting window, we fit eq.\,(\ref{eq:powerlaw}) in the temperature range of $T_{\rm min}\,{\leq}\,T\,{\leq}\,T_{\rm max}$, setting the lowest temperature at $T_{\rm min}\,{=}\,0.36$ K and varying the highest temperature $T_{\rm max}$ from 1.87 to 0.82 K. We actually observed a systematic decrease in the exponent $\phi$ with decreasing $T_{\rm max}$, as shown in Fig.~2 in paper I. 
\begin{figure}[t]
\includegraphics[width=8.5cm, clip]{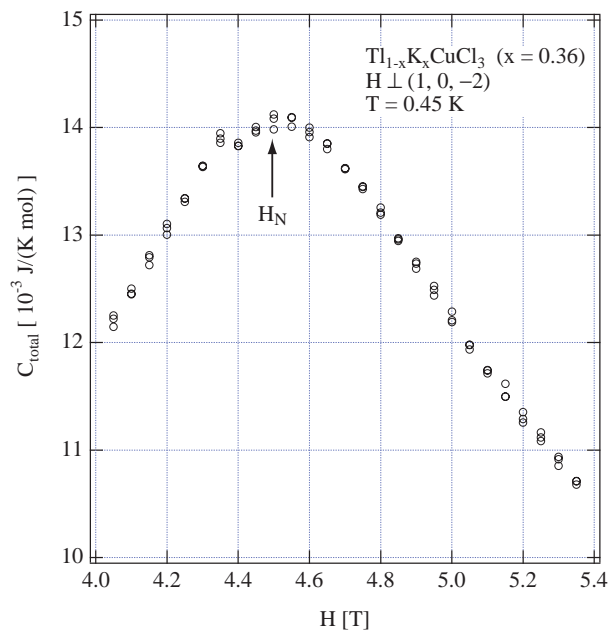}
\caption{Field scan of specific heat in Tl$_{1-x}$K$_x$CuCl$_3$ with $x\,{=}\,0.36$ measured at 0.45\,K. The arrow denotes the transition field $H_{\rm N}(T)$. }
\label{fig1}
\end{figure}

\begin{figure}[t]
\includegraphics[width=8.5cm, clip]{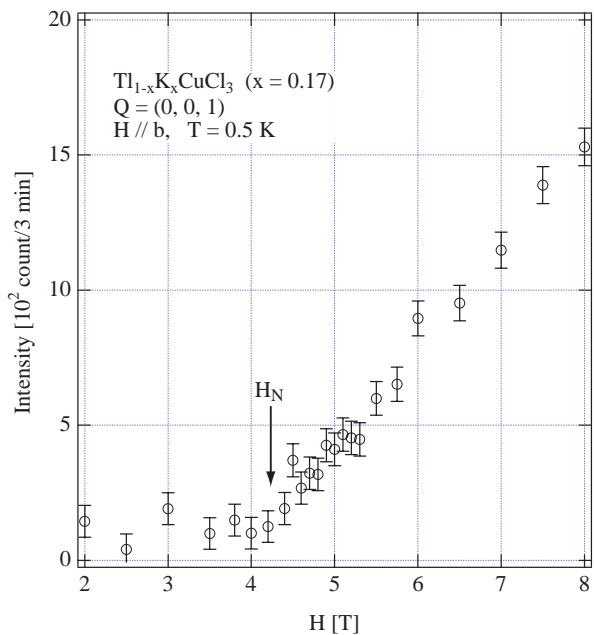}
\caption{Magnetic field dependence of magnetic peak intensity in Tl$_{1-x}$K$_x$CuCl$_3$ with $x\,{=}\,0.17$ for $(0, 0, 1)$ measured at 0.5 K.}
\label{fig2}
\end{figure}

Comment authors claim that the phase boundaries for both $x\,{=}\,0$ and $x\,{\neq}\,0$ can also be expressed by the same critical exponent ${\phi}\,{=}\,1$. It is obvious from Fig.~1(c) in paper I that the phase boundary for pure TlCuCl$_3$ is perpendicular to the field axis, while those for $x\,{\neq}\,0$ are not. Because of large three-dimensional interdimer interactions of the order of 1\,meV, magnetic excitations in TlCuCl$_3$ are largely dispersive in three dimensions~\cite{Cavadini,Oosawa,Matsumoto}. This leads to a small triplon mass and a small coefficient $A$ in eq.\,(\ref{eq:powerlaw})~\cite{Nikuni,Misguich}. For this reason and a large saturation field of $H_{\rm s}\,{\simeq}\,90$\,T, the temperature-field region of $T\,{<}\,2$\,K and $H-H_{\rm c}\,{<}\,1$\,T can be considered as a critical region for pure TlCuCl$_3$, which is close to the quantum critical point (QCP). Near the QCP, the transition field $H(T)_{\rm N}$ does not exhibit a large temperature dependence.

Because for pure TlCuCl$_3$, the transition field $H_{\rm N}(T)$ below 1 K scarcely depends on temperature and the error in determining $H_{\rm N}(T)$ from the field scan of specific heat is ${\pm}\,0.05$\,T, it is insufficient to obtain a correct exponent only from data points below 1 K. For this reason, we used data points up to 2 K and obtained ${\phi}\,{=}\,1.53$. This analysis should be appropriate. 

Actually, the specific heat peak shown in Fig.~1 is rounded. This should be ascribed to the instrumental resolution. Comment authors strongly suspect that the field-induced transition in Tl$_{1-x}$K$_x$CuCl$_3$ is smeared by an antisymmetric interaction, such as the Zeeman term with the staggered $g$ tensor and the DM interaction, and thus, the transition is not a true phase transition. Indeed, it is difficult to assume that there is no antisymmetric interaction in Tl$_{1-x}$K$_x$CuCl$_3$, but we can estimate the upper limit of the magnitude of the antisymmetric interaction. 
The magnitude of the anisotropy can be evaluated from the ESR linewidth ${\Delta}H$, which is given by the magnetic anisotropy that does not commute to the total spin~\cite{VV}. Near 0.5 K, the ESR line shape is between Lorentzian and Gaussian, which indicates that exchange narrowing is less effective. In such a case, the ESR linewidth corresponds to the magnitude of the local field due to magnetic anisotropy~\cite{VV}. From ${\Delta}H\,{\simeq}\,0.1$\,T near 0.5 K, the total energy ${\Delta}E$ of the anisotropy including the staggered Zeeman term and the DM interaction can be evaluated as ${\Delta}E\,{\simeq}\,0.01$\,meV. The magnitude ${\Delta}E_{\rm AS}$ of the antisymmetric interaction is smaller than ${\Delta}E$, and thus, ${\Delta}E_{\rm AS}$ is much smaller than the intradimer interaction of 5.7\,meV and the interdimer interactions of the order of 1\,meV~\cite{Cavadini,Oosawa,Matsumoto}. Therefore, we consider that the effect of the antisymmetric interaction on the field-induced transition is negligible.

Neutron diffraction data may be useful for determining whether the transition is well-defined. At present, we have diffraction data only for $x\,{=}\,0.17$. In Fig.~2, we show the field dependence of the $(0, 0, 1)$ magnetic reflection for $x\,{=}\,0.17$ measured at 0.5 K in magnetic fields parallel to the $b$ axis. A clear bend anomaly due to the field-induced phase transition is observed at $H_{\rm N}\,{=}\,4.2$\,T. The sharpness of the bend anomaly for $x\,{=}\,0.17$ is similar to that in the case of pure TlCuCl$_3$~\cite{Tanaka}. 
   
Comment authors claim that specific heat diverges at a continuous transition point. However, this seems to be a misunderstanding on their part. It is established that the critical exponent ${\alpha}$ for specific heat is negative for the three-dimensional $XY$ and Heisenberg universality classes, and thus, the specific heat does not diverge at the transition points for these cases~\cite{Guillou}.

\end{document}